\documentclass[conference]{IEEEtran}
\IEEEoverridecommandlockouts
\usepackage{cite}
\usepackage{amsmath,amssymb,amsfonts}
\usepackage{algorithmic}
\usepackage{graphicx}
\usepackage{textcomp}
\usepackage{xcolor}
\usepackage{hyperref}
\usepackage{pgfplotstable}
\pgfplotsset{compat=1.14}
\usepackage{listings}

\def\BibTeX{{\rm B\kern-.05em{\sc i\kern-.025em b}\kern-.08em
    T\kern-.1667em\lower.7ex\hbox{E}\kern-.125emX}}
    
\definecolor{blind_safe_one_scheme_three_colors}{RGB}{102,194,165}
\definecolor{blind_safe_two_scheme_three_colors}{RGB}{252,141,98}
\definecolor{blind_safe_three_scheme_three_colors}{RGB}{141,160,203}
\definecolor{blind_safe_one_scheme_four_colors}{RGB}{166,206,227}
\definecolor{blind_safe_two_scheme_four_colors}{RGB}{31,120,180} 
\definecolor{blind_safe_three_scheme_four_colors}{RGB}{178,223,138}
\definecolor{blind_safe_four_scheme_four_colors}{RGB}{51,160,44}
\definecolor{blind_safe_one_scheme_seven_colors}{RGB}{118,42,131}
\definecolor{blind_safe_two_scheme_seven_colors}{RGB}{175,141,195}
\definecolor{blind_safe_three_scheme_seven_colors}{RGB}{231,212,232}
\definecolor{blind_safe_four_scheme_seven_colors}{RGB}{247,247,247}
\definecolor{blind_safe_five_scheme_seven_colors}{RGB}{217,240,211}
\definecolor{blind_safe_six_scheme_seven_colors}{RGB}{127,191,123}
\definecolor{blind_safe_seven_scheme_seven_colors}{RGB}{27,120,55}

\definecolor{butter1}{rgb}{0.988,0.914,0.310}
\definecolor{butter2}{rgb}{0.929,0.831,0.000}
\definecolor{butter3}{rgb}{0.769,0.627,0.000}

\definecolor{orange1}{rgb}{0.988,0.686,0.243}
\definecolor{orange2}{rgb}{0.961,0.475,0.000}
\definecolor{orange3}{rgb}{0.808,0.361,0.000}

\definecolor{chocolate1}{rgb}{0.914,0.725,0.431}
\definecolor{chocolate2}{rgb}{0.757,0.490,0.067}
\definecolor{chocolate3}{rgb}{0.561,0.349,0.008}

\definecolor{chameleon1}{rgb}{0.541,0.886,0.204}
\definecolor{chameleon2}{rgb}{0.451,0.824,0.086}
\definecolor{chameleon3}{rgb}{0.306,0.604,0.024}

\definecolor{skyblue1}{rgb}{0.447,0.624,0.812}
\definecolor{skyblue2}{rgb}{0.204,0.396,0.643}
\definecolor{skyblue3}{rgb}{0.125,0.290,0.529}

\definecolor{plum1}{rgb}{0.678,0.498,0.659}
\definecolor{plum2}{rgb}{0.459,0.314,0.482}
\definecolor{plum3}{rgb}{0.361,0.208,0.400}

\definecolor{scarletred1}{rgb}{0.937,0.161,0.161}
\definecolor{scarletred2}{rgb}{0.800,0.000,0.000}
\definecolor{scarletred3}{rgb}{0.643,0.000,0.000}

\definecolor{aluminium1}{rgb}{0.933,0.933,0.925}
\definecolor{aluminium2}{rgb}{0.827,0.843,0.812}
\definecolor{aluminium3}{rgb}{0.729,0.741,0.714}
\definecolor{aluminium4}{rgb}{0.533,0.541,0.522}
\definecolor{aluminium5}{rgb}{0.333,0.341,0.325}
\definecolor{aluminium6}{rgb}{0.180,0.204,0.212}

\definecolor{codegreen}{rgb}{0,0.6,0}
\definecolor{codegray}{rgb}{0,0,0}
\definecolor{codepurple}{rgb}{0.58,0,0.82}
\definecolor{backcolour}{rgb}{0.95,0.95,0.92}

\lstdefinestyle{mystyle}{
    backgroundcolor=\color{backcolour},   
    commentstyle=\color{codegreen},
    keywordstyle=\color{magenta},
    numberstyle=\tiny\color{codegray},
    stringstyle=\color{codepurple},
    basicstyle=\ttfamily\footnotesize,
    breakatwhitespace=false,         
    breaklines=true,                 
    captionpos=b,                    
    keepspaces=true,                 
    numbers=left,                    
    numbersep=5pt,                  
    showspaces=false,                
    showstringspaces=false,
    showtabs=false,                  
    tabsize=2
}
\begin{document}

\title{TDO-CIM: Transparent Detection and Offloading for Computation In-memory}

\author{Kanishkan Vadivel$^{1}$/Lorenzo Chelini$^{1,2}$, Ali BanaGozar$^{1}$, Gagandeep Singh$^{1,2}$, Stefano Corda$^{1,2}$, Roel Jordans$^{1}$, \\ Henk Corporaal$^{1}$
\\ $^1$Eindhoven University of Technology, $^2$ IBM Research Zürich \\\{k.vadivel, l.chelini\}@tue.nl}
 
\newboolean{showcomments}
\setboolean{showcomments}{true}
\ifthenelse{\boolean{showcomments}}
{ \newcommand{\mynote}[3]{
     \fbox{\bfseries\sffamily\scriptsize#1}
        {\small$\blacktriangleright$\textsf{\emph{\color{#3}{#2}}}$\blacktriangleleft$}}}
{ \newcommand{\mynote}[4]{}}
\newcommand{\vk}[1]{\mynote{KV}{#1}{red}}
\newcommand{\lz}[1]{\mynote{LZ}{#1}{green}}
\newcommand{\Ali}[1]{\mynote{Ali}{#1}{orange}
}
\newcommand{\gaganCheck}[1]{\mynote{Gagan}{#1}{blue}
}

\maketitle
\begin{abstract}


Computation in-memory is a promising non-von Neumann approach aiming at completely diminishing the data transfer to and from the memory subsystem. Although a lot of architectures have been proposed, compiler support for such architectures is still lagging behind. In this paper, we \textit{close} this gap by proposing an end-to-end compilation flow for in-memory computing based on the LLVM compiler infrastructure. Starting from sequential code, our approach automatically detects, optimizes, and offloads kernels suitable for in-memory acceleration. We demonstrate our compiler tool-flow on the PolyBench/C benchmark suite and evaluate the benefits of our proposed in-memory architecture simulated in Gem5 by comparing it with a state-of-the-art von Neumann architecture.
\end{abstract}

\begin{IEEEkeywords}
LLVM, compute in memory, memristor, pattern matching, Polly, Loop Tactics
\end{IEEEkeywords}

\section{Introduction}
As we are moving toward exascale computing, the memory wall~\cite{Wulf:1995:HMW:216585.216588} is becoming one of the toughest challenges for the traditional von Neumann architecture. Not only does the cost of moving data dwarf the cost of a floating-point operation but also the memory bandwidth is not able to meet the demand of today's applications~\cite{Singh2019}. Consequently, new architectures with a radical departure from the traditional von Neumann architecture start to arise. Computing in-memory (CIM) is one of them. CIM aims at processing information and storing computation data on the same physical unit using emerging devices referred to as memristive devices. Memristive devices such as phase change memory devices (PCM) can store data within their conductance state, which can be changed by altering the amorphous/crystalline phase within the device~\cite{le2018mixed}. Computation, on the other hand, is carried out through various physical mechanisms such as Ohm's and Kirchhoff's laws. Memristor devices are organized in a computational memory unit---which we refer to as CIM tile. As storing and processing happen in the same physical device, CIM completely diminishes the overhead of data movement between the CPU and the main memory, enabling data-intensive task in an efficient manner~\cite{8450603}. A huge body of research has been done on the architecture side~\cite{7446049,7551380,7544414}. However, before CIM can be established as the de-facto solution for HPC and IoT applications a \textit{leap forward need to be done on the compilation toolchain and software stack}, which is the purpose of this paper. Our contributions are:
    \begin{itemize}
        \item An \textit{end-to-end compilation flow} for CIM devices, which allows to \textit{automatically} and \textit{transparently} invoke CIM acceleration, without any user intervention. Therefore, enabling legacy code to exploit in-memory acceleration.
        \item A \textit{lightweight run-time library} for data allocation, transfer and execution of computational tasks on the CIM device.
        \item We \textit{evaluate the benefits of CIM computation} in terms of energy and performance by comparing it with a current state-of-the-art von Neumann architecture using the PolyBench/C benchmark suite. 
    \end{itemize}

\section{Compute In Memory Architecture}
In this section, we first briefly discuss the physic behind the PCM device (Section~\ref{sub:memristor_physic}). Afterward, we show how such devices can be interconnected together in a crossbar-like structure (Section~\ref{sub:cim_tile}) to create the basic block of our accelerator (Section~\ref{sub:cim_acc}).
\begin{figure}
  \begin{center}
    \includegraphics[width=8cm]{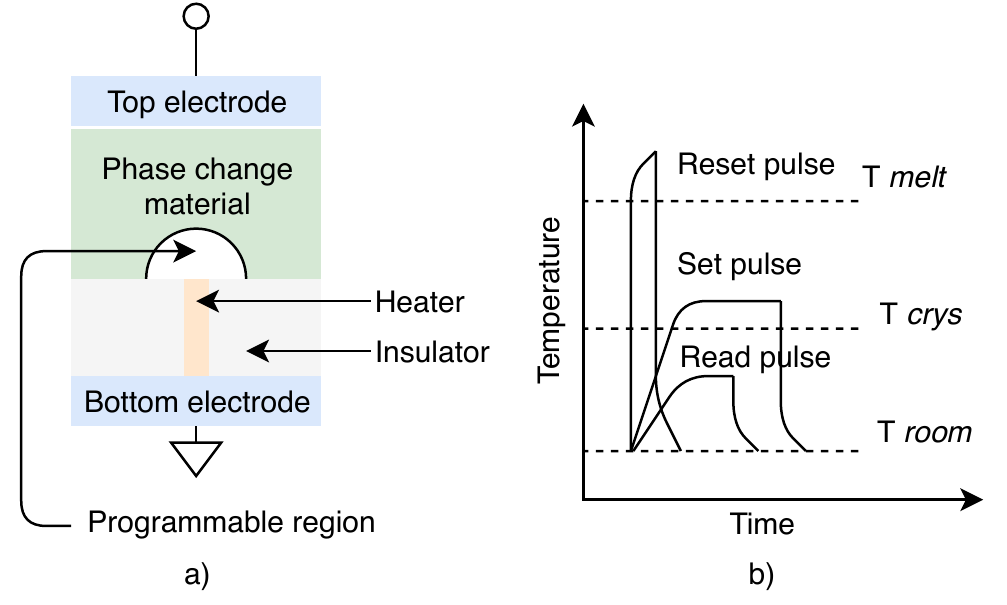}
    \caption{Cross section of a PCM device (a) and its programming pulses (b).}
    \label{fig:pcm}
  \end{center}
\end{figure}
\begin{figure*}
  \begin{center}
    \includegraphics[width=18cm]{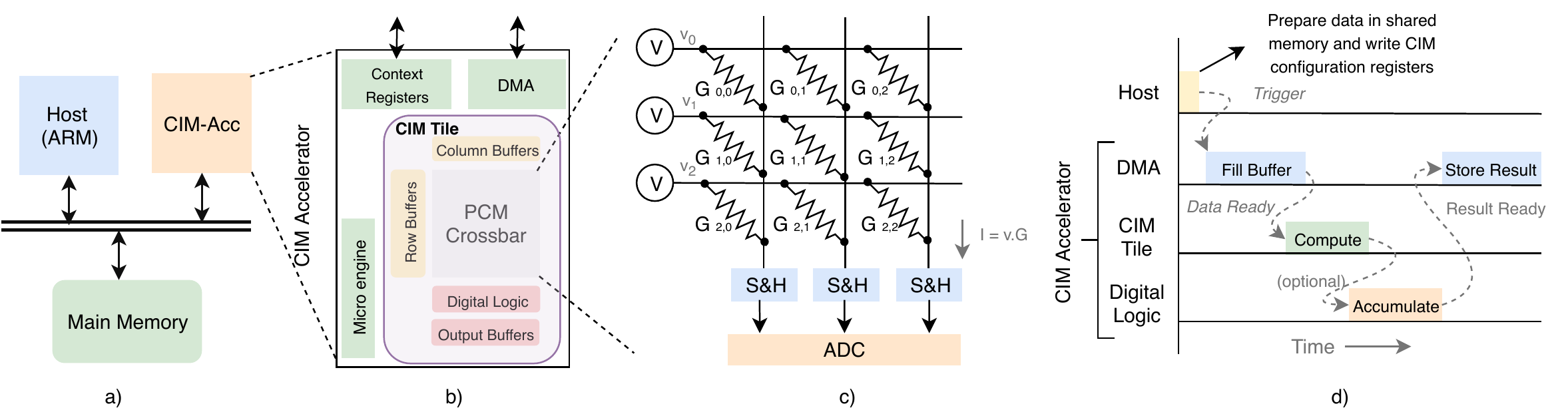}
    \caption{Overview of the emulated system (a). A more detailed view of our developed CIM accelerator (b). A memristor-based crossbar (c). Timeline of a kernel execution on our CIM accelerator (d).}
    \label{fig:cim_arch}
  \end{center}
\end{figure*}
\subsection{Memristor Basics}
\label{sub:memristor_physic}

PCM is a type of non-volatile memory that stores information by changing the cell resistance, switching between amorphous and crystalline states. The transition between the two states happens as a consequence of the application of external voltages that exceed the threshold voltage of a device. Figure~\ref{fig:pcm} (a) shows a cross-section of a PCM device. The phase change material is sandwiched between two electrodes, and current is applied through the heater in order to change the material state. A short, but intense, pulse---known as reset pulse---is used to bring the material in the amorphous phase (high-resistance). Contrarily, to switch back to low resistance the set pulse---a lower and longer pulse---is applied. To read the device, an even lower pulse (read pulse) is used (Figure~\ref{fig:pcm} (b)). Thanks to the excellent scaling capabilities of PCM devices---which allows increasing main memory capacity in a cost-effective and power-efficient way---it is expected that PCM will play a significant role in future memory architectures~\cite{raoux2014phase}. But before this can happen, one main challenge needs to be addressed: endurance. PCM devices can stand $10^6$ - $10^8$ writes before they wear out making the lifetime of a PCM-based system last for a few years~\cite{Qureshi:2009:ELS:1669112.1669117}. Despite a lot of effort on architecture support for wear-leveling and smart algorithms for data re-placement, no prior work tries to address this endurance problem at compile time. TDO-CIM addresses this obstacle by revisiting two common compiler transformations: tiling and fusion (Section~\ref{sub:specific_opt}).


\subsection{CIM Tile}
\label{sub:cim_tile}
The electrical conductance/resistance of the PCM depends on the material phase of the device. A single PCM can achieve several resistance levels that can be exploited for in-memory computations~\cite{emerging_mem}. Each resistance level can be used to encode a particular binary value. For instance, a PCM device with $2^M$ levels can support a maximum of $M$-bit computation at full-resolution. To support higher resolutions with a low precision device, multiple columns in a crossbar can be used~\cite{fpca}. Each column in crossbar computes partial results, and the final result is computed by a weighted sum using traditional CMOS technology. Figure~\ref{fig:cim_arch} (c) shows how PCM devices can be organized in a crossbar-like structure to execute matrix-vector multiplication. A matrix can be stored in the crossbar as the conductance state of the PCM devices ($G_x,_y$ values). Afterward, the input vector is fed as a set of voltages to the crossbar, which multiplies by the conductance values. The resulting current sensed at the columns is the analog dot-product result~\cite{brain-inspired}. The output currents are converted back into digital signals by analog to digital converters (ADC). To further improve the energy efficiency, ADCs are shared amongst multiple columns which are reused using sample and holds (S\&H)~\cite{ISAAC}. In addition to the previously mentioned analog components, a digital interface is required to hook the CIM tile (Figure~\ref{fig:cim_arch} (b)) with traditional CMOS logic. The digital interface is composed of row/column buffers, output buffers, and a digital logic block. The row/column buffers act as a data and mask registers for the crossbar~\cite{cim-sim}. During write operation, the column buffers contain the data that has to be written on the crossbar, and the row buffers supply a row-enable signal. Similarly, during a compute operation, the column buffers supply column-enable signal and the row buffers latch the inputs. The computed result will be stored in the output buffers. The digital logic block implements scalar compute functionality (i.e., reduction functions) to perform post-processing on the crossbar result.

\subsection{Accelerator Organization}
\label{sub:cim_acc}
A CIM tile, a micro-engine, and a DMA unit for load and store operations make a standalone accelerator. The core is the CIM tile which computes a standard matrix-vector multiplication (GEMV) of complexity $O(N^2)$ in $O(1)$ constant time. The matrix-matrix computation (GEMM) can be implemented as a series of matrix-vector operations, and therefore the accelerator supports both GEMV and GEMM. 

The accelerator uses a shared global memory interface for data sharing and exposes a set of context registers to the system via a memory-mapped IO interface. Context registers are used for control and offloading, and are read or written by the host. The micro-engine translates the high level-parameters stored in the context registers into a series of circuit-level operations such as loading the data from shared memory to row/column buffers, configuring the mask values, triggering the computation on CIM tile, and writing back the results from the output buffers to the shared memory. Additionally, it manages the control flow involved in decomposing GEMM to a series of GEMVs and supports double buffering for all the registers in the accelerator to hide the data latency of the memory accesses. Figure~\ref{fig:cim_arch} (d) shows a timeline of the events that happen after the host trigger the CIM accelerator.

\begin{figure}
  \begin{center}
    \includegraphics[width=9cm]{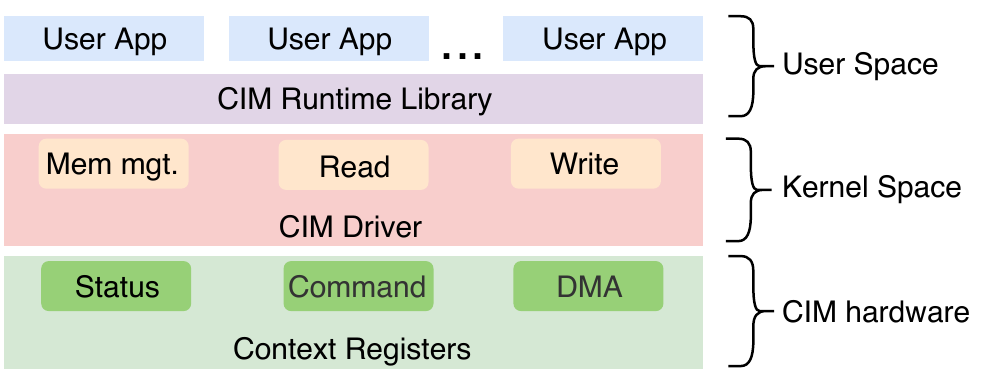}
    \caption{Overview of the CIM's software stack.}
    \label{fig:sw_stack}
  \end{center}
\end{figure}

\subsection{Hardware Model}
Figure~\ref{fig:cim_arch} (a) shows our emulated system with a host, main memory, and a CIM accelerator connected through a system bus. We implement the CIM accelerator as a cycle-accurate model integrated into the Gem5 simulator~\cite{gem5}. The accelerator is based on a port-mapped IO (PMIO) and a DMA interface. The PMIO interface exposes context registers to the system, and the DMA provides a memory interface to the accelerator. The host mimics a dual-core Arm-A7 processor based on~\cite{gem5_config}. For the experiments in the paper, we run the simulator in full-system mode to capture the effects of the operating system, device drivers, and hardware interactions.


\subsection{Software Stack}
A software stack (Figure~\ref{fig:sw_stack}) allows applications running in the user space to interact directly with the hardware. The software stack is divided into kernel-space and user-space. At the lowest level of the stack, the kernel-space CIM driver reads and writes to the context registers of the accelerator through a \texttt{ioctl} system call. Besides, the driver translates the virtual address used by the host processor to a physical address as the accelerator can work only with physical addresses. On the other hand, the user-space CIM API is responsible for encoding CIM runtime library calls into context register parameters.
Furthermore, with the help of the CIM driver, it implements the support for allocating and releasing the physically-contiguous pages in shared memory via the contiguous memory allocator (CMA) APIs exposed by the Linux kernel~\cite{kernel_cma}. The use of CMA offers two main benefits compared to the traditional malloc-based approach: 1) the size of the shared memory region is not limited by the page boundary; 2) there is no need for explicit memory management in the driver routines, which diminishes overhead in the host. 

To enforce memory coherence in the shared memory region, the kernel driver triggers a cache flush on the host side before invoking the accelerator. The accelerator, on his part, uses only un-cachable requests for memory access which automatically enforces memory coherence. Once the accelerator completes its execution, it updates the status in a specific context register. The host can either wait on spinlock or continue with other tasks and check the status of such register periodically. 

\section{Overview of the CIM compiler}
\label{sub:compiler_overview}
\begin{figure*}
  \begin{center}
    \includegraphics[width=18cm]{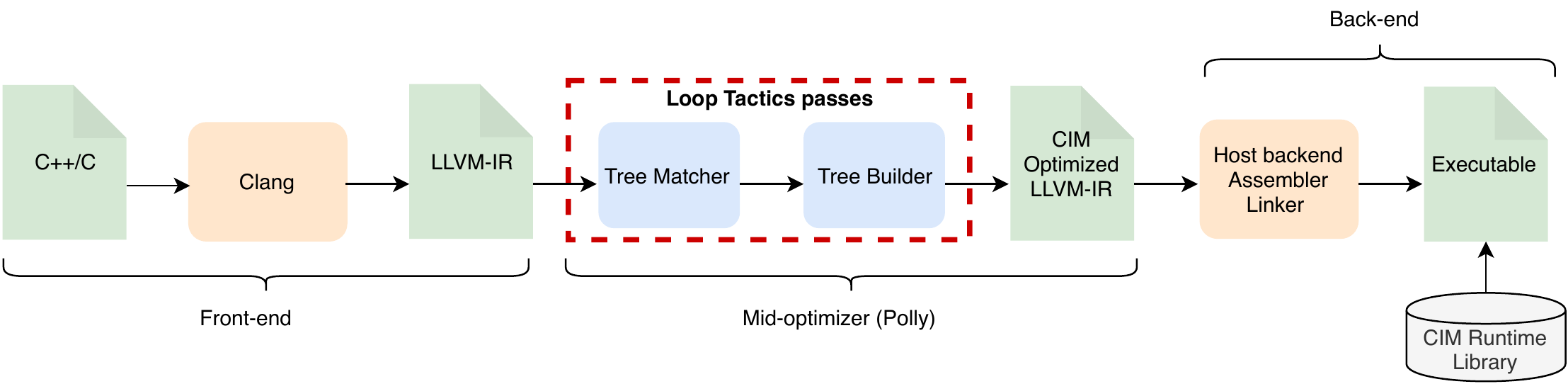}
    \caption{LLVM-based compilation flow developed for the CIM accelerator.}
    \label{fig:flow}
  \end{center}
  \vspace{-5mm}
\end{figure*}

The high-level design of our compilation flow is shown in Figure~\ref{fig:flow}. It follows a classical compiler design with a front-end, a mid-level optimizer and targets specific back-ends. We extend this flow by introducing 1) Loop Tactics~\cite{zinenko2018declarative, myTaco}---a state-of-the-art declarative optimizer---in the mid-level. Loop Tactics enables automatic detection and offload of specific computational patterns; 2) A lightweight runtime library that provides optimized performance and memory usage for the CIM device. The library has been designed to be used directly by the application programmer, or an optimizer (i.e., Loop Tactics). It exposes a host-callable C API, similar to what cuBLAS or MKL offers for Nvidia GPU and Intel CPU, respectively. 
\subsection{A Bird's-eye View of TDO-CIM}
The entry point in our compilation flow is an application written in a high-level language (i.e., C++). To handle a variety of languages front-ends lower the high-level language to an intermediate representation (IR) on which all the subsequent optimizations are spelled. For our work we use the LLVM compiler infrastructure, hence adopting its intermediate representation LLVM-IR. Given an application, we can use any of the LLVM-based front-ends (i.e., Clang) to lower the high-level language to LLVM-IR. At LLVM-IR level we rely on the polyhedral optimizer Polly~\cite{grosser2012polly} to detect, extract and model compute kernels. Internally Polly represents the schedule of each detected kernel as a tree, which we refer to as schedule tree. Schedule tree~\cite{verdoolaege2014schedule} is the way of representing the execution strategy of each kernel by mapping each dynamic statement instance with its execution order. This mapping is implicitly defined by the node parent-child relation within the tree. Loop optimizations and device mapping are expressed as tree modifications and carried out by Loop Tactics, which works as additional passes within Polly. Loop Tactics' passes consume schedule trees and output a CIM-optimized schedule. The modified tree is then passed back to Polly which lowers it back to an imperative AST and then further down to LLVM-IR. In the back-end LLVM-IR is lowered to final executable. It is at this stage of the compilation pipeline where we link our executable with the CIM runtime library. Listing~\ref{lst:code_generation} shows at the top a GEMM kernel in C++ code, while at the bottom the mid-level optimizer output as pseudo-C++. For our example, we assume single-precision operands. The GEMM kernel has been detected and swapped by Loop Tactics with a function call to the CIM runtime library (\texttt{polly\_cimBlasSGemm}). Blas parameters (i.e., values of alpha or leading dimensions) are automatically collected or computed by Loop Tactics. In addition, Loop Tactics inserts an initialization call to configure the CIM hardware (\texttt{polly\_cimInit}) as well as all the function calls to orchestrate the data movement to and from the device (i.e., \texttt{polly\_cimMalloc} and \texttt{polly\_cimDevToHost}). 
 
\vspace{5pt}
\begin{lstlisting}[language=C]
for (int i = 0; i < M; ++i)
  for (int j = 0; j < N; ++j) {
  C[i][j] = beta * C[i][j];
    for (int k = 0; k < K; ++k)
      C[i][j] += alpha * A[i][k] * B[k][j];
  }
\end{lstlisting}
 
\vspace{1.50mm}
\begin{lstlisting}[language=C, label={lst:code_generation}, caption={High-level code for a generalize matrix multiplication (GEMM) kernel (top). Loop Tactics generated code to offload GEMM kernel to the CIM accelerator (bottom).}]
/* ... */
// initialize CIM device 0
polly_cimInit(0);
// allocate data on CIM device
polly_cimMalloc((void**)&cim_C, M*N*4);
polly_cimMalloc((void**)&cim_A, M*K*4);
polly_cimMalloc((void**)&cim_B, K*N*4);
// execute GEMM kernel on CIM device
polly_cimBlasSGemm(transA, transB, M, N, K,
  &alpha, cim_A, lda, cim_B, ldb, 
  &beta, cim_C, ldc);
// copy C back to host
polly_cimDevToHost(cim_C, host_C, M*N*4);
\end{lstlisting}
\vspace{5pt}

  
  

\subsection{TDO-CIM specific optimizations}
\label{sub:specific_opt}
We revisit loop fusion and tiling in the light of this new CIM computing paradigm trying to minimize write operations to crossbar to enhance endurance.
\paragraph*{Revisited Loop Fusion}
Loop fusion is a performance-oriented transformation that combines two loop nests in a new single-loop nest. In our case, we focus on a specific case of loop fusion: kernel fusion. Consider two consecutive kernels X and Y, with Y following X directly. We fuse X and Y if both kernels have the same access patterns (i.e., both are GEMM kernels) and are independent. Two kernels are independent if Y doesn't read from or write to any output of X, and Y does not write to any input of X. An example is shown in Listing~\ref{lst:kernel_fusion}.

\newcommand{\cfbox}[2]{%
    \colorlet{currentcolor}{.}%
    {\color{#1}%
    \fbox{\color{currentcolor}#2}}%
}


\vspace{5pt}
\begin{lstlisting}[language=C, label={lst:kernel_fusion}, caption={Independent kernels with shared input (A matrix). TDO-CIM exploits shared inputs to increase endurance by avoiding multiple writes on the memristor crossbar.}]
for (int i = 0; i < M; ++i)
  for (int j = 0; j < N; ++j) {
    for (int k = 0; k < K; ++k)
s1:   C[i][j] += @\ul{A[i][k]}@ * B[k][j];
  }
for (int i = 0; i < M; ++i)
  for (int j = 0; j < N; ++j) {
    for (int k = 0; k < K; ++k)
s2:   D[i][j] += @\ul{A[i][k]}@ * E[k][j];
  }
\end{lstlisting}
\vspace{5pt}

\begin{figure}
\pgfplotsset{compat = 1.3}
\pgfplotsset{major grid style={dotted,aluminium2!50!black}}
\begin{tikzpicture}[]
\begin{axis}[
    /pgf/number format/.cd,
    scale = 0.5,
    use comma,
    1000 sep={},
    width=0.9\textwidth,
    height=0.4\textwidth,
    xlabel={PCM cell endurance (number of writes in million)},
    ylabel={System lifetime (years)},
    xmin=0, xmax=45,
    ymin=0, ymax=50,
    xtick={10, 20, 30, 40, 50},
    ytick={0, 8, 16, 24, 32, 40, 48},
    legend style={at={(0.02,1.02)},anchor=north west},
    legend columns=1, 
    legend style={draw=none, fill=none},
    every axis plot/.append style={ultra thick},
    ymajorgrids=true,
    grid style=dashed,
    legend cell align={left},
]

\addplot[
    color=blind_safe_one_scheme_four_colors,
    mark=square,
    ]
    coordinates {
    (10, 6) (15, 9) (20, 12) (25, 15) (30, 18) (35, 21) (40, 24)
    };
    \addlegendentry{Naive mapping}

\addplot[
    color=blind_safe_three_scheme_four_colors,
    mark=*,
    dotted,
    mark options={solid},
    smooth
    ]
    coordinates {
    (10,12) (15,18) (20, 24) (25, 30) (30, 36) (35, 42) (40, 48)
    };
    \addlegendentry{``Smart'' mapping}
\end{axis}
\end{tikzpicture}
\caption{Impact of TDO-CIM fusion transformation for the code in Listing~\ref{lst:kernel_fusion}.}
\label{fig:endurance_fusion}
\vspace{-5mm}
\end{figure}
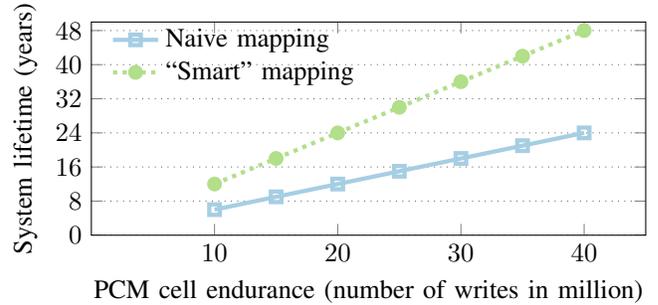

By fusing two kernels, we get the following advantages: 1) we reduce the number of calls to the runtime library by using batched operations. The GEMMs in Listing~\ref{lst:kernel_fusion} will be replaced by a single \texttt{polly\_cimBlasGemmBatched} instead of two calls to \texttt{polly\_cimBlasSGemm}. The interface for the batched operation is similar to the one provided for \texttt{polly\_cimBlasSGemm} with the only exception of having arrays of pointers instead of single pointers. 2) We increase endurance by exploiting possible shared inputs. The A matrix (Listing~\ref{lst:kernel_fusion}) is shared and remains constant; we exploit this by writing only A in the crossbar and streaming B and E from the input buffers. This allows writing only one matrix on the crossbar in contrast with a naive mapping where B end E would have been written, and A streamed from the input buffers. Figure~\ref{fig:endurance_fusion} shows the expected lifetime for the PCM crossbar comparing the naive mapping and the ``smart'' mapping applied by TDO-CIM. The $x$-axis shows the PCM cell endurance in an interval between 10 million to 40 million writes which is in the expected lifetime interval of a PCM device ($10^6$ to $10^8$). The expected lifetime is computed by applying the following equation~\cite{Qureshi:2009:ELS:1669112.1669117}:
\begin{equation}
    SystemLifeTime = \frac{CellEndurance * S}{B}
\end{equation}
where $S$ is the crossbar size, 512 KB in our case, while $B$ is the write traffic in GB/s for the kernel in Listing~\ref{lst:kernel_fusion}. $B$ is obtained by diving the total number of writes by the kernel execution time.  We assume squared matrices of 4096 byte-elements and the writes to be localized uniformly across the entire crossbar. As can be seen from Figure~\ref{fig:endurance_fusion} the ``smart'' mapping allows to improve endurance by a factor of 2.

\paragraph*{Revisited Tiling Transformation}
Let us now focus on our revisited tiling optimization. Tiling is a well-known transformation to improve locality by reducing the reuse distance of memory accesses to the same location. Consider statement s1 in Listing~\ref{lst:kernel_fusion} and assume that matrix A doesn't fit in the CIM crossbar. We use tiling to split A into multiple tiles such that the working set of a single tile fits in the CIM crossbar. We then apply loop interchange on the tile loops \texttt{jj} and \texttt{kk} such that we can reuse tiles of A in \textit{consecutive} executions of the point loops, hence once more increasing endurance. The outcome of our tiling transformation is shown in Listing~\ref{lst:kernel_tiling}, where the point loops will be replaced by a call to \texttt{polly\_cimBlasSGemm}.


\begin{lstlisting}[language=C, label={lst:kernel_tiling}, caption={Loop tiling and interchange to fit the operand on the CIM crossbar and reduce the number of writes by reusing A tiles in consecutive execution of the point loops.}]
// tile loops
for (int ii = 0; ii < SIZE; ii += TILE)
  for (int kk = 0; kk < SIZE; kk += TILE)
    for (int jj = 0; jj < SIZE; jj += TILE) 
      // point loops
      for (int i = ii; i < ii + TILE; i++)
        for (int j = jj; j < jj + TILE; j++)
          for (int k = kk; k < kk + TILE; k++) 
            C[i][j] += A[i][k] * B[k][j]; 
\end{lstlisting}

\section{Demonstration and Evaluation}
\label{sub:demons}
In this section, we quantify the benefits of CIM computation for a set of linear-algebra kernels from the Polybench/C benchmark suite compiled with TDO-CIM. 

\paragraph{Experimental Setup}
\label{sub:exp_setup}
We use the system shown in Figure~\ref{fig:cim_arch} (a). We select an energy efficient dual-core Arm-A7 with a shared L2 cache. The simulator is a cycle-accurate model that imitates the functionality of the memristor computations and surrounding digital blocks~\cite{cim-sim}. The memristor crossbar is an 8-bit 256x256 PCM crossbar based on IBM's 4-bit PCM~\cite{8450603}. To mimic an 8-bit cell with a 4-bit cell, two adjacent columns are used, one for 4 MSBs and the other for 4 LSBs. The final result is computed by a weighted sum of MSB and LSB columns in the digital logic block. The energy and latency model for the crossbar and mixed-signal circuitry is from~\cite{8450603} and \cite{ISAAC} respectively. The energy model for the rest of the digital blocks is based on a synthesis report of commercial 40nm finFET technology. Table~\ref{tab:systemchar} summarises our system configuration and energy model.

\begin{table}[]
\caption{CIM and Host System Configuration.}
\label{tab:systemchar}
\centering
\scalebox{0.98}{
\begin{tabular}{l|l}
\hline
\textbf{CIM Parameter}& \textbf{Value}\\
\hline
\textbf{PCM Crossbar} \\
Technology(256x256 @8-bit)&  IBM PCM 2x(256x256 @4-bit)  \\
Compute and Write Latency/8-bit & 1$\mu$s and 2.5$\mu$s  \\
Compute Energy/8-bit & 200fJ (2x 100fJ/4-bit PCM) \\
Write Energy/8-bit & 200pJ (2x 100pJ/4-bit PCM) \\ 
\hline
Energy for Mixed signal circuit & 3.9nJ (@1.2GHz) \\
Input/Output buffer Energy (1.5KB) & 5.4pJ/byte-access\\
Digital Logic & 40pJ/GEMV for weighted sum\\
& and 2.11pJ/extra ALU operation \\
Energy for DMA and microEngine & \textless0.78nJ \\
\hline
\hline
\textbf{Host CPU Spec}  \\
\hline
2xArm-A7 @1.2GHz & 2GB LDDDR3 @933MHz \\
L1-I/D-32KB, L2-2MB & 128pJ/inst\footnotemark (including cache)\\
\hline
\end{tabular}
}
\vspace{-5mm}
\end{table}
\footnotetext{Based on \textit{Ara: Energy-Efficient RISC-V}, Matheus et al. 2019.}

\begin{figure*}[h]
	\centering
	\includegraphics[ width=\columnwidth]{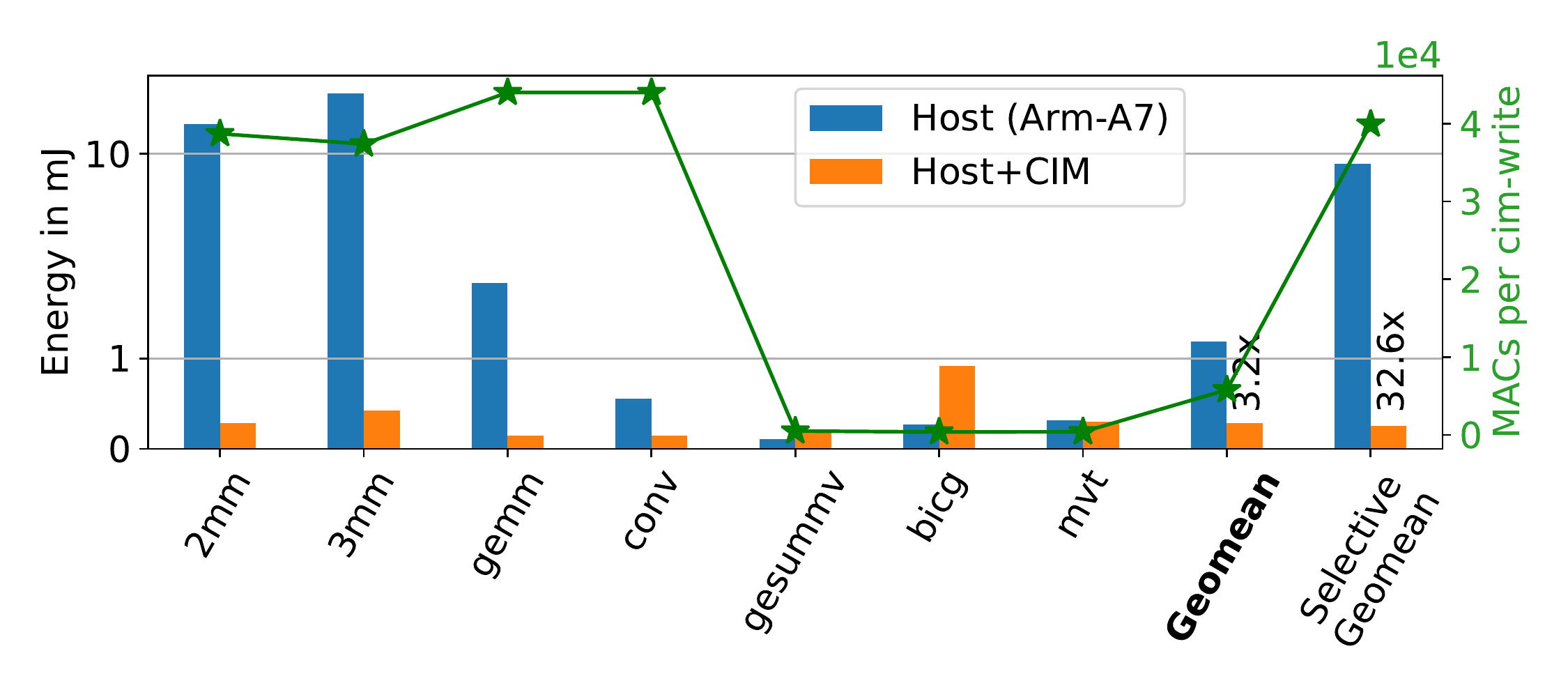}	    \includegraphics[width=\columnwidth]{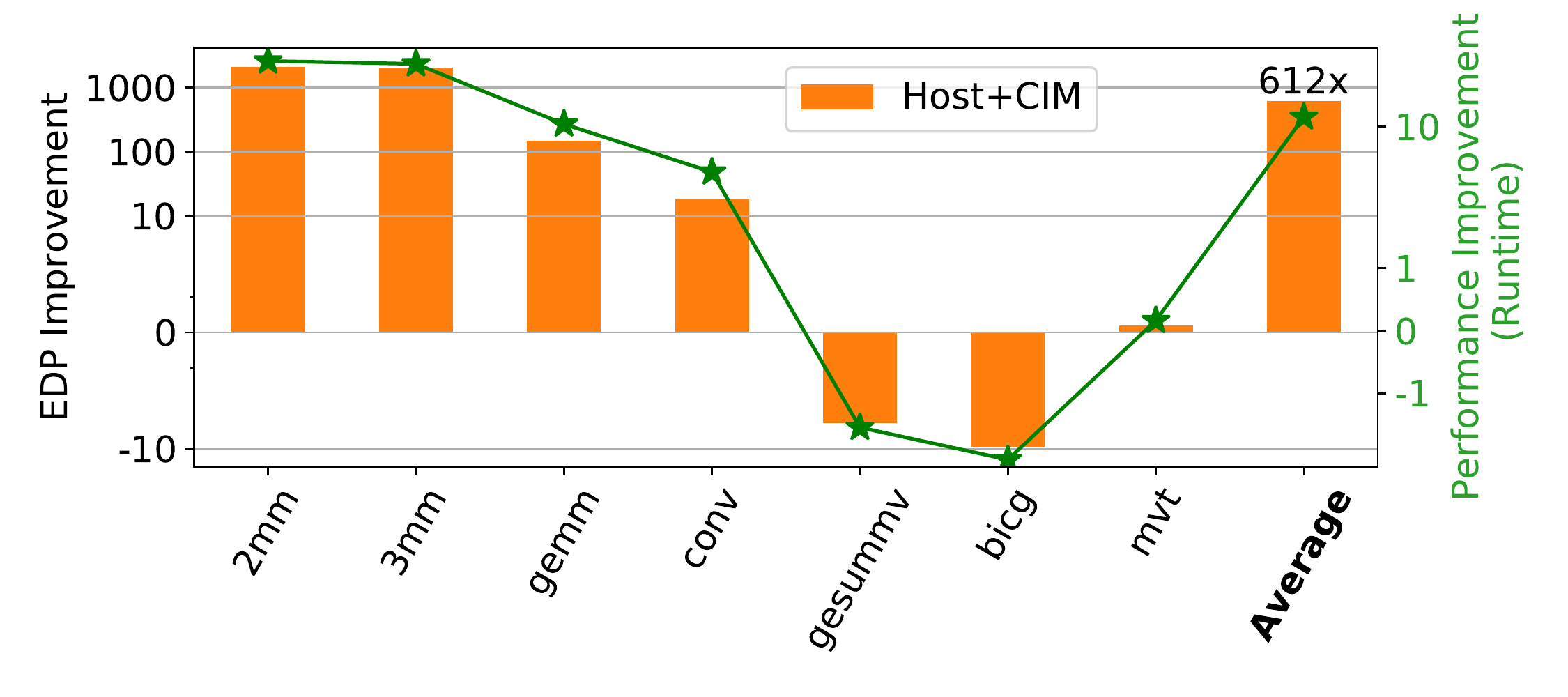}
	\caption{Energy (left) and Energy-delay-product (right) improvement for CIM computation.}
	\label{fig:cim_results}	
	\vspace{-5mm}
\end{figure*}

\paragraph{Performance Evaluation}
We use the compilation string shown in footnote\footnote{\texttt{clang -O3 -march-native \\ clang -O3 -march-native -enable-loop-tactics}} for the host and the host+CIM, respectively. Dynamic instruction count and run-time are profiled in Gem5 by inserting ROI markers. For energy estimates, we use the numbers shown in Table~\ref{tab:systemchar}. We do not include DRAM energy numbers in the estimates as the host and CIM-accelerator generate the same amount of traffic by accessing the same data. Figure~\ref{fig:cim_results} (left) shows the energy numbers obtained for the reference platform (Arm-A7), and for the Arm-A7+CIM where the kernel execution is performed on the in-memory accelerator. For the host, the energy numbers include the energy spent on computation and in the memory hierarchy. For the CIM, the energy numbers incorporate the energy spent on the driver (host side) and in the accelerator. GEMM-like kernels: \texttt{2mm}, \texttt{3mm}, \texttt{gemm}, and \texttt{conv} were able to achieve good energy improvements over the reference system. This is not the case for GEMV-like kernels (\texttt{bicg}, \texttt{mvt}, \texttt{gesummv}) due to their low compute intensity. From the CIM perspective, the compute intensity for a given kernel can be formulated as $\frac{Number-of-MAC-operations}{Number-of-CIM-writes}$ which is very low for GEMV-like kernels as can be seen in Figure~\ref{fig:cim_results} (left). With such low compute intensity the energy is dominated by the overhead in host for offloading computations to accelerator and the number of writes which are costly for the CIM device 200pJ/byte (see Table~\ref{tab:systemchar}). Figure~\ref{fig:cim_results} (right) shows the energy-delay-product (EDP). It follows the same trend as the energy plot. We gain for GEMM-like kernels (up to 612x) while we lose for GEMV-like.

\section{Related Work}
\paragraph*{Code offloading}
Several works address the issue, TOM~\cite{7551394} being perhaps the very first of them. TOM proposes an offloading decision based on a simple cost function. The idea is to statically identify the code section with the highest potential in bandwidth saving. Similarly, Pattanik~et.~al. propose an affinity prediction model based on memory-related metrics to decide where a given kernel should be executed (i.e., main CPU or in-memory accelerator)~\cite{7756764}. Previously mentioned works target GPU as an in-memory accelerator. On the other hand, in our case, we are targeting a memristor crossbar which means that only specific kernels must be offloaded as the accelerator is capable of executing only GEMM and GEMVs-like kernels. Nair~et.~al. propose a code offloading based on OpenMP 4.0 user annotation~\cite{7095154}. Contrary, our approach is completely transparent to the application and does not require any user intervention to exploit CIM acceleration. CAIRO relies on an LLC cache profiler and analytical models to decide potential offloading candidates~\cite{Hadidi:2017:CCT:3154814.3155287}. The LLC profiler is not integrated into the compilation flow and requires to characterize the behavior of each kernel offline. Other works expose CIM acceleration via API~\cite{ 7551380, 7544414}, which requires significant changes in the application, reducing application readiness, and hurdling widespread adoption.

\paragraph*{Enhance PCM endurance}
Software and hardware wear-leveling techniques to distribute write operations uniformly across the memory module have bee studied extensively. Hardware techniques require additional storage tables to keep track of heavily written blocks, that will be periodically get remapped to the lowest wear-out ones~\cite{Qureshi:2009:ELS:1669112.1669117}. Software techniques, on the other hand, rely on lazy write-back policy~\cite{Qureshi:2009:ELS:1669112.1669117}, dynamic data management~\cite{5763127}, data migration and recomputation~\cite{Hu:2010:RWA:1837274.1837363}. All previous approaches are orthogonal to TDO-CIM, which tries to enhance endurance at compile time by intelligently mapping array references to the CIM crossbar.

\section{Conclusion}
We present an end-to-end compilation flow for in-memory computing. Our approach automatically identifies, optimizes, and offloads computing kernels to our in-memory accelerator. We compile a set of linear-algebra kernels from the Polybench/C benchmark suit and prove the benefits of in-memory computation by comparing our in-memory architecture simulated in Gem5 with a state-of-the-art von Neumann architecture. The results show the benefits of in-memory computing by achieving average energy reduction of 32.6x and energy-delay-product improvement of 612x. We expect our compiler and Gem5 emulator to boost researches in the field by providing a transparent and automatic flow to compile entire applications on the CIM architecture and perform domains-space exploration by tweaking our simulator.



\section*{Acknowledgments}
This research is supported by EC Horizon 2020 Research and Innovation Program through MNEMOSENE project under Grant 780215 and the NeMeCo grant agreement, id.\,676240.


\end{document}